# Phase-only transmissive spatial light modulator based on tunable dielectric metasurface

*Shi-Qiang Li, Xuewu Xu, Rasna Maruthiyodan Veetil, Vytautas Valuckas, Ramón Paniagua-Domínguez, and Arseniy I. Kuznetsov\**

Institute of Materials Research and Engineering, A*STAR (Agency for Science, Technology and Research), 138634, Singapore

*Corresponding author, email: arseniy_kuznetsov@imre.a-star.edu.sg

**Abstract**

Rapidly developing augmented reality (AR) and 3D holographic display technologies require spatial light modulators (SLM) with high resolution and viewing angle to be able to satisfy increasing customer demands. Currently available SLMs, as well as their performance, are limited by their large pixel sizes of the order of several micrometres. Further pixel size miniaturization has been stagnant due to the persistent challenge to reduce the inter-pixel crosstalk associated with the liquid crystal (LC) cell thickness, which has to be large enough to accumulate the required 2π phase difference. Here, we propose a concept of tunable dielectric metasurfaces modulated by a liquid crystal environment, which can provide abrupt phase change and uncouple the phase accumulation from the LC cell thickness, ultimately enabling the pixel size miniaturization. We present a proof-of-concept metasurface-based SLM device, configured to generate active beam steering with >35% efficiency and large beam deflection angle of 11°, with LC cell thickness of only 1.5 μm, much smaller than conventional devices. We achieve the pixel size of 1.14 μm corresponding to the image resolution of 877 lp/mm, which is 30 times larger comparing to the presently available commercial SLM devices. High resolution and viewing angle of the metasurface-based SLMs opens up a new path to the next generation of near-eye AR and 3D holographic display technologies.

## Introduction

Spatial light modulators (SLMs) have widespread applications ranging from 3D video projection[1] and additive manufacturing[2] to quantum[3] and adaptive optics.[4] The most versatile type of SLMs are phase-only SLMs, which are capable of reconfiguring the phase retardation of light transmitted through or reflected from each pixel arbitrarily and independently without changing the light intensity. Ideally, in order to obtain the best performance for a phase-only SLM, each pixel in the SLM should be able to provide a phase shift coverage of at least $2\pi$ radians. Typically, SLMs achieve this by using liquid crystals (LC) as the active medium, which is a liquid like material constituted by anisotropic molecules, called LC directors. These directors tend to align with each other, which thus endues the LC medium with a large uniaxial anisotropy of refractive index. The difference between extraordinary refractive index ($n_e$), when light polarization is parallel to the directors, and ordinary one ($n_o$), when it is perpendicular, $\Delta n = n_e - n_o$ typically ranges from 0.2 to 0.4 in the visible spectral range. These directors, moreover, rotate under the influence of an applied electric field as they have induced dipole moments, providing means to dynamically control the refractive index along a given direction.

One of the limitations of current LC-based SLMs is their slow modulation/switching speed. They operate at a typical modulation frequency from 60 to 480 Hz for the nematic LCs and from 2 to 32.6 kHz for the ferroelectric ones.[5] This is sufficient for most of projection and display applications but can hardly be used for applications like Light Detection and Ranging (LiDAR). The modulation speed could be increased with the reduction of the thickness of the LC layer ($h_{LC}$), as the time constant of LC director rotation is proportional to $h_{LC}$ to the power of two.[6] However, in order to attain a phase modulation range of $2\pi$ in a transmissive or reflective device, $h_{LC}$ cannot be smaller than $\lambda/\Delta n$ or $\lambda/2\Delta n$ respectively (where $\lambda$ is the wavelength of operation), which corresponds to the thickness of several microns for visible spectral range.

Another limitation of LC-based SLMs is their large pixel size, which is above 3 microns for the best reflective SLM devices. The reason for this limitation is that downsizing the pixels further while keeping the required thickness of LC layer introduces a cross-talk between the pixels, owing to the broadened phase profile and high level of detrimental fringing fields.[6-10] This, in turn, limits the field of view (FOV) of the SLM, which is defined as the angular coverage of the first diffractive order, inversely proportional to the pixel size, and represents one of the most important figure of merits (FOMs) for phase-only SLMs. This problem is more severe for transmissive SLMs, which have a smaller filling factor due to the space occupied by transistors and an LC thickness that is twice larger than that of reflective SLMs to obtain the same phase accumulation. As an example, a typical commercial transmissive SLM (such as HOLOEYE LC 2012) has a pixel pitch of 36 microns, which gives a maximum FOV of 0.7° only. These constraints of existing phase-only SLMs severely limit their application areas.

Recently, a new class of flat optical components, so called metasurfaces, relying on completely different mechanisms to engineer the wavefronts of light, caught great attention in the research community. These systems abruptly modify the optical phase of incoming light by designated amounts using sub-diffractive optical elements, so-called nanoantennas.[11-19] To do so, several different mechanisms were proposed, such as using the Pancharatnam-Berry phase[20-21] or using resonances excited in the nanoantennas (either plasmonic or Mie-type[22]). Since they were first proposed, they have found many potential applications, including meta-lenses to replace bulky lens systems relying on phase accumulation along the direction that light propagates,[16-17, 19, 23-24] and generation of complex beams or holography,[13, [18] just to mention some. While these are static optical components, for which the phase coverage can be achieved via arranging the constituent nanoantennas with different shapes, sizes, or orientations, for various applications it is also important to find ways to adaptively tune the phase brought to light by the nanoantennas during usage.[25-35] Ultimately, if each single nanoantenna element of a metasurface could be dynamically tuned by applying electrical voltages then it would act as an SLM device with a pixel size smaller than the wavelength of light.

In this work, we propose and demonstrate an idea on how to achieve this by integrating the nanoantennas into an LC-SLM device. In this case, the main phase accumulation happens inside the nanoantennas rather than in the LC layer itself, thus uncoupling it from the LC layer thickness. At the same time, switching of the LC orientation around the nanoantennas changes their local environment, modifying the spectral position of the nanoantenna resonances and opening a door to dynamically control the phase retardation experienced by an incoming light beam. This helps to reduce the thickness of the LC layer, which is crucial to solve the aforementioned limitations of these devices, while not requiring significant modifications to the existing SLM fabrication technology.

First results on integration of nanoantennas inside LC cells[25-26, 28, 36-38] have shown the possibility to obtain the spectral shift of nanoantenna resonances and associated amplitude modulation of light using electrically[38] and thermally[36-37] induced LC directors switching. Recent results have also demonstrated switchable beam deflection at a fixed angle with an LC-integrated gradient metasurface.[28] However, a universal active device which is capable of arbitrary beam manipulation, such as 3D holographic animation or beam steering at multiple angles, has not been demonstrated so far. This type of devices requires uniform pixels which can be switched individually to give rise to the desired phase front, as discussed above.

Here, we use the large refractive index change introduced by the electrically modulated LC layer to manipulate the Mie resonances of titanium dioxide nanoantennas embedded within. We show that integration of nanoantennas on top of transmissive SLM pixels may enable them to work with good efficiency with just a 1.5 micron thick LC layer and 1.14 micron pixel size (less than twice the operational wavelength of 660 nm). To prove our concept, we demonstrate an efficient one-dimensional (1-D) 3-phase-level transmissive SLM device. It enables the realization of a beam deflection function with a maximum beam steering angle of 11° with respect to the normal, corresponding to a FOV of 22°, and a diffraction efficiency of 36% with respect to the incident light, significantly outperforming not only the state-of-art transmissive LC-SLMs but also solid-state ones, which typically have small steering angles of a few degrees only.[6, 39-40]

## Background and Simulations

For a transmissive phase-only nanoantenna SLM one need to design each nanoantenna pixel to transmit light efficiently and be able to cover the whole $2\pi$ phase range. One possible way to achieve this is using the so-called Huygens' metasurface concept, realized here via dielectric nanoantennas supporting both electric and magnetic dipole resonaces.[15, 41-42] When these two modes are simultaneously excited and overlapped, a full-range phase shift from 0 to $2\pi$ around the resonance peak can be obtained.[41] In addition, following Kerker's theoretical analysis,[43] the overlapped and balanced magnetic and electric dipole resonances lead to a total cancellation of backward scattering, resulting in a close-to-unity transmission through the array.[14-15, 44]

To achieve high efficiencies in the visible spectral range we design nanoantennas made of $TiO_2$,[24, 45] which has negligible absorption at the visible wavelengths as well as a relatively high refractive index (n ~2.5) sufficient to obtain resonances inside a liquid crystal environment (taken here to be the E7® from Merck, $n_o$ ~ 1.5 and $n_e$ ~ 1.7, see Supplementary Information (SI) Figure S1). The spectral positions and strengths of magnetic and electric dipoles are strongly dependent on the aspect ratio of the nanoantennas[41] as well as on the periodicity of the array (p).[46] In order to find the optimum dimensions, we perform full-wave finite-element-method simulations using COMSOL Multiphysics software to gain an understanding of the influence of the key parameters. We start by considering a square lattice of cylindrical $TiO_2$ nanoantennas embedded in the LC with thickness $h_{LC}$ = 1500 nm. The LC is sandwiched between two glass plates and its director is oriented in-plane, i.e. $\theta_{LC}$ = 0° as shown in Figure 1a, corresponding to the so-called nematic phase of the LC. The incident light wave impinges normally onto the device surface with its electric field polarized parallel to the LC director, the wave thus experiencing the extraordinary refractive index of the LC. In Figure 1b&c, we show the maps of transmission magnitude and phase profile for different radii of the $TiO_2$ nanoantennas sweeping around the optimized dimensions in the vicinity of targeted wavelength of 660 nm while keeping a fixed height of 205 nm. The period in all cases is p = 360 nm, so that the unit cell is sub-diffractive. One can see that a high transmission and a full $2\pi$ phase coverage can be achieved in a wide range of wavelengths from 650 to 675 nm indicating a high parameter tolerance of our nanoantenna design. This high tolerance is optimized for the chosen nanoantenna height and becomes narrower at different heights as shown in the Supplementary Information, Figures S2 and S3.

It should be noticed, however, that there is a strong dispersion of phase around the resonance, making it rather sensitive to the radius of the nanoantennas. Thus, the size uniformity needs to be well preserved as, otherwise, it may affect the performance of the device. For example, in Figure 1c, it can be seen that a mere 20 nm radius variation covers the entire $2\pi$ phase range at a wavelength of 660 nm.

With the optimized nanoantenna dimensions, we study the phase coverage of the transmitted light through the nanoantenna array when the LC director orientation is switched from in-plane to out-of-plane. This kind of LC rotation can practically be achieved by applying an electrical bias between the top and bottom electrodes sandwiching a nematic LC with pre-aligned, in-plane parallel directors.[47] The rotation of LC directors has two effects. Firstly, it modifies the permittivity tensor of the LC layer, modulating the refracting index of the environment around the nanoantennas and thus changing the spectral position and intensity of their resonances.[28, 36, 38] Consequently, the phase induced by the nanoantenna resonances at a particular wavelength changes accordingly. Secondly, the modified permittivity tensor also induces a difference in the phase accumulation in the LC layer itself.

In Figure 2(a), we plot the calculated phase retardation of light incident normally to the optimized nanoantenna array (h = 205 nm, R = 135 nm, p = 360 nm, $h_{LC}$ = 1500 nm) across the spectrum of interest. We plot the results for LC directors lying in the x-z plane with $\theta_{LC}$ varying from 0° to 90° with 5° increment, different color markers representing different LC director orientations. The size of the markers is proportional to the intensity of transmitted light (the size of the markers used in the legend corresponding to 50% transmission). We see that above 670 nm, the relative phase difference for the device at different LC orientations does not exceed $0.8\pi$, which is not enough to provide $2\pi$ phase coverage required for SLMs. This is the spectral region in which the nanoantennas are not resonant with the incident light and the phase retardation is entirely attributed to the phase accumulation in the LC layer.

Below 670 nm, the resonances excited in the nanoantennas generate strong phase variations. In Figure 2(a), we have highlighted three series of results corresponding to LC director orientations of 0°, 45° and 90°. The phase differences between these three series is evenly spaced in the wavelength region between 660 nm and 670 nm with approximate phase difference of around $2\pi/3$ to each other. Furthermore, they also have similar transmittance ranging between 60% and 90% thus satisfying the two basic criteria for a 3-phase level spatial light modulator – high transmission and evenly spaced phase coverage.

To demonstrate a proof of concept device functionality of the 3-phase-level spatial light modulator, we simulate light beam bending induced by a gradient phase profile introduced by neighboring nanoantennas with the three different LC orientations having equally spaced phase retardation levels. First, from the simulation we obtained the electric field distribution inside the LC cell at the wavelength of 665 nm for the three LC orientations separately and put them side-by-side in Figure 2(b). One can see that the wave-fronts of the incident plane wave are only slightly affected by the propagation through the LC layer for the three cases. This indicates little phase difference accumulated inside the LC cell itself. During the interaction with nanoantennas though, the wave-fronts of transmitted light experience a different phase retardation with respect to each other, forming three adjacent phase steps. As a consequence, in the far field, light transmitted through a system consisting of a periodic repetition of these three unit cells should be deflected out of the normal to the surface, in a direction perpendicular to the tilted wave front indicated by light dotted lines in Figure 2(b).

We then simulate a real device structure consisting of cylindrical $TiO_2$ nanoantennas placed on ITO electrodes patterned on top of a glass substrate, as schematically shown in Figure 3(a). The system is covered with another ITO-coated glass, serving as a common, ground electrode, and the LC is sandwiched between the substrate and the cover, embedding the nanoantennas. Instead of one nanoantenna for each phase level, we choose to have three of them on each bottom electrode. This is to reduce the effect of phase reset at the edge of each unit cell, which may cause phase profile broadening deviating it from the desired phase-front. In addition, this gives enough width to the electrodes to accommodate the fringing electric field brought about by the finite LC domain wall thickness (i.e. when the LC domain on one electrode orients in a different direction to the domain on the adjacent electrodes), which is on the order of hundred nanometers.[48] According to our simulations, further increase of the number of particles per pixel does not significantly improve the performance (see Table S1 in SI). The LC orientation above the three neighboring electrodes is put at 0°, 45° and 90° respectively, corresponding to the respective phase differences of 0, $2\pi/3$ and $4\pi/3$ radians, reproducing the gradient phase profile (similar to Figure 2(b)) to deflect the incident beam. This structure is then repeated periodically as a supercell of an infinite gradient metasurface. The beam deflection efficiency of the proposed SLM device numerically calculated using a commercial Finite-Difference-Time-Domain (FDTD) solver (Lumerical Inc.) is plotted in Figure 2(c). It can be observed that, at approximately 665 nm, there is a clear increase of the transmitted power into the -1 diffraction order, together with a strong reduction of the transmission into the 0 and +1 orders. This translates into the predicted beam deflection, consistent with the analysis of homogeneous arrays shown in Figure 2(a) and 2(b).

## Experimental Results

To experimentally realize the proof-of-concept beam deflecting SLM device, rectangular 20 nm thick ITO electrodes are fabricated on top of a glass substrate. The length and width of each electrode are 100 μm and 1.08 μm, respectively, and their spacing is 60 nm. Then 3 rows of cylindrical $TiO_2$ nanoantennas with a height of 205 nm, diameter of 270 nm and period of 360 nm are placed on top of each of the electrodes, as shown in Figure 3(a). The electrodes with nanoantennas are then embedded into a 1.5 μm thick LC cell with another ITO electrode on top. The top electrode is covered by an LC alignment layer with preferred alignment direction perpendicular to the long side of the bottom electrodes. Detailed nanofabrication and LC cell assembling procedures are described in Methods section in Supplementary Information. Before the LC cell assembly, the dimensions of the fabricated nanostructures have been verified through scanning electron microscopy (SEM) and reflectance spectral measurement (see Figure S4). In operation, two of each three neighboring bottom electrodes are connected to the two terminals of an external voltage source. One of the connected electrodes and the top electrode are put to the ground state and the other is biased to induce an out-of-plane switching of the LC directors above the electrode (left panel in Figure 3(a)). The third electrode is left floating/unconnected to pick up the electric potential in between the biased and the grounded electrodes, so that the LC directors can attain an intermediate level of rotation leading to deflection of the incoming light beam. Note that incident light beam can also be deflected to the other direction by switching the applied voltages, as shown in the right panel in Figure 3(a). In the case of no bias applied to the electrodes, light beam simply passes through the device with exactly the same phase at all three electrodes, as shown in the central panel in Figure 3(a).

SEM images of the fabricated device before the liquid crystal infiltration are shown in Figures 3(b) and 3(c). A typical unit cell, consisting of three electrodes, is highlighted by a yellow-shaded box. There is an obvious contrast in the SEM images between the unconnected electrode (the one at the bottom part of the yellow box) and the other two electrodes. The gaps between the electrodes, which appear darker in the SEM images, can also be clearly observed in the magnified views in the right panels of Figures 3(b) and (c). The SEM image contrast in both cases is related to the higher conductivity of the connected ITO electrodes compared to the floating ones.

Optical performance of the device is characterized using the spectrally-resolved back focal plane imaging technique[23, 49-50] (see Methods section in SI for more details). Figures 4(a) and (b) show the beam deflection results at the wavelength of 663 nm under different biases. At zero bias, electrodes are in a homogeneous LC environment and the incident light goes straight into the zeroth order (situation depicted in Figure 3(a) middle panel). When we start to ramp up the electrical bias applied to the set of electrodes on the left side (Figure 3(a), left panel) from 0V to values above 6V, the beam starts to get deflected into the first diffraction orders. With relatively low voltages (see Figure 4(a) at 6V and 8V), the zeroth order intensity decreases and both -1 and +1 diffraction orders increase simultaneously. This implies that the unconnected central electrode has not yet gained enough potential at small bias voltages. Thus, the LC at the central electrode retains the same orientation as the grounded electrode on the right. In this case, the device functions as a symmetric grating splitting the beam to both positive and negative diffraction orders. With further increase of the bias on the left electrode, the LC directors on top of the left electrode saturate in the out-of-plane position keeping it for all subsequent higher voltage values. Nevertheless, for increasing bias, the central electrode picks up an increasing electric potential pulled up by the left electrode. This rotates the LC directors on top of the central electrode and the phase front starts to tilt. As a result, the beam deflects based on the 3-phase level wave front described in Figure 2(b). We found that the highest efficiency of beam deflection (at 11° angle) occurs at 12 V, as shown in Figure 4(a). For higher bias levels, the electric field and, consequently, LC rotation at the unconnected electrode raises above the optimum value and the efficiency drops (see SI Figure S5).

We also test the beam deflection into the opposite direction by reverting the configuration, i.e. grounding the top and bottom left electrodes and biasing the bottom right electrodes (as shown schematically in Figure 3(a), right panel). The results are shown in Figure 4(b). As expected, in this case the beam at an optimum voltage is deflected to the right, a reversed situation as compared to the case with the left electrode biased. It should be noted, however, that now the zeroth order is more suppressed than in the preceding case and the optimum deflection occurs at a slightly higher voltage. We attribute this slight difference to imperfections of the fabricated device. The central electrode is slightly closer to one side of the connected electrodes, which may be the origin of the observed asymmetry in the beam deflection efficiency between the left and right configurations.

In Figure 4(c) and (d), we plot the measured diffraction efficiency spectra of the device (i.e. the transmitted power normalized by the incident one) into the three main diffraction orders, -1, 0 and +1 (denoted as $T_{-1}$, $T_0$ and $T_{+1}$), at the optimum beam deflection conditions for each case (namely left and right deflection configurations). The overall trend agrees well with the simulated results plotted in Figure 2(c), despite the efficiency obtained from the experiment being slightly lower. In the best case, the experimental beam deflection efficiency reaches 36% of the incident power channeled into the -1 diffraction order at 663 nm compared to ~48% obtained from the simulations (Figure 2(c)). One should note that there are several effects that are not considered in the FDTD simulations, such as the fringing fields at the gaps between the electrodes, size non-uniformity of the nanoantennas, and imperfections in the LC director alignment, which all may contribute to this slight discrepancy. We expect the efficiency to get higher with improvement of the fidelity in the nanofabrication, better LC alignment, and external biasing of the center electrode. The latter would help, moreover, to reduce the voltage required to obtain the beam deflection, as no pulling of the floating electrode would be necessary.

It is also important to note that while in our proof of concept experiment only 3 angular positions (-11°, 0°, +11°) for beam deflection have been demonstrated, the proposed 3-phase level transmissive SLM is capable to deflect the incoming beam at multiple different angles within the maximum demonstrated range of 22°. This will require a proper electrical connection and addressing of each ITO electrode individually.

## Conclusions

We demonstrated a 1D phase-only nanoantenna-based transmissive spatial light modulator (SLM) device, which has a good efficiency of 36% with a pixel size of only ~1micron and a large viewing angle span of 22°. The device works based on an abrupt phase accumulation provided by a tunable Huygens' metasurface consisting of $TiO_2$ dielectric nanoantennas embedded in a liquid crystal cell. The phase accumulation is dynamically modulated by changing the nanoantenna environment via liquid crystal director rotation through application of external bias. The presence of the nanoantennas allows a reduction of the liquid crystal layer thickness required to achieve the necessary phase modulation by more than a half compared to traditional SLMs. This, in turn, alleviates the phase broadening effect and the fringing field effect, which are the main limiting factors of the traditional SLM devices, allowing significant reduction of the pixel size. This concept can further be extended to 2-dimensional phase-only liquid crystals SLMs, opening up the possibility of making ultra-high-resolution devices which may perform faster and have a larger field of view with high efficiency.

## Figure Captions

Figure 1. Optimization of the nanoantenna geometry through simulations. (a) Schematic of the simulated unit cell. A $TiO_2$ nanodisk with a height $h_a$ and radius R is placed on top of a glass substrate. The disk is embedded in a liquid crystal (E7) environment with a thickness $h_{LC}$ = 1.5 micron. The liquid crystal directors are initially aligned parallel to the X-axis, while the light is incident normal to the surface of the substrate (along the Z-axis) and is polarized along the X-axis. A square array of the nanoantennas is formed by translating the unit cell in the both X and Y directions with a period p = 360 nm. (b) Zero order transmission spectrum of the array for nanoantenna radius sweeping from 130 nm to 150 nm and fixed height of 205 nm. (c) Spectrum of the phase of the transmitted wave for nanoantennas with a fixed height of 205 nm and varying radius, sweeping from 130 nm to 150 nm. The refractive indices of $TiO_2$ and E7 are plotted in SI (Figure S1).

Figure 2. Simulation results for the device under different LC director orientations. (a) Calculated relative phase retardation of the wave front passing through the nanoantenna unit cell (with h = 205 nm, R = 135 nm, and p = 360 nm). Color of the markers changes gradually from blue to orange to yellow-green mapping the change of orientation of LC directors from in-plane (pointing to X-direction, 0°) to out-of-plane (pointing to Z-direction, 90°). The size of the markers is proportional to the transmittance calculated with respect to the transmission through the LC layer in the absence of the nanoantennas. The size of the markers in the legend corresponds to 50% transmittance. The series of markers highlighted are those that are used to achieve the optimum 3-phase level SLM device. (b) Electric field distribution ($E_X$) in the unit cell under three different LC director orientations at the wavelength of 665 nm. The excitation light is polarized along the X-axis and impinges from the top to the bottom at a normal incidence to the substrate surface (indicated by the green solid line as a guide to eyes). The light-yellow dotted lines outline the phase fronts which would emerge from the device after being retarded differently in adjacent cells having different orientations of LC directors. The normal to these lines represents the direction of the outgoing deflected beam. The slight misalignment of phase and intensity contrast above the top boundary between glass and liquid crystal is due to interference between a small amount of reflected wave with the incident wave. (c) Calculated spectral transmission into the three main diffraction orders: -1, 0 and +1, denoted by $T_{-1}$, $T_0$ and $T_{+1}$, respectively. The beam deflection into -1 order peaks at around 665 nm, consistent with the analysis in (a) and (b).

Figure 3. Metasurface-based SLM: principle of operation and device structure. (a) Schematic drawings of the SLM and its operation concept. The LC molecules (E7) are pre-aligned uniformly along the short axis of the electrodes and parallel to the surface of the substrate. When all electrodes are at the same potential, the LC layer is uniformly aligned in-plane and the incident beam travels straight through the device, as shown in the center panel. When the left electrode is biased, the directors of the LC layer on top the electrode are rotated, pointing out-of-plane. The bias difference between the left and right electrodes also induces an intermediate bias at the unconnected/floating electrode rotating the LC directors on top of it to an intermediate state. This creates a tilted phase front as shown in Figure 2(b), causing the beam to be deflected to the left, as depicted in the left panel. Conversely, biasing the opposite (right) electrode causes the beam to be deflected to the other direction (right) as shown on the right panel. (b) Top view scanning electron microscope (SEM) images showing the fabricated device structure. The observed bright stripes are the unconnected/floating

electrodes inducing charging effect leading to the bright contrast. A zoomed-in view is shown on the right panel. (c) Titled SEM images of the whole device structure. Each device covers a region of 120 by 100 microns. The zoomed-in view of individual electrodes is shown in the right panel. Scale Bars:  Left panel in (b): 5 microns; Right panel in (b): 600 nm; Left panel in (c): 15 microns; Right panel in (c): 250 nm. The false color, yellow shaded areas highlight a unit cell of the device.

Figure 4. Performance of the fabricated device. (a) Measured far field beam deflection under the bias voltage applied to the bottom left electrode while keeping the top electrode and the bottom right electrode grounded. (b) Measured far field beam deflection under the bias voltage applied to the bottom right electrode while keeping the top electrode and the bottom left electrode grounded. In both cases, the first five diffraction orders are shown (from -2 to +2) at the wavelength of 663 nm. (c) Measured transmission diffraction efficiency (transmission intensity normalized to incident one) for the three main diffraction orders (from -1 to +1) corresponding to 12V bias in (a). (d) Measured transmission diffraction efficiency for the three main diffraction orders corresponding to 13V bias in (b).

Figure 1

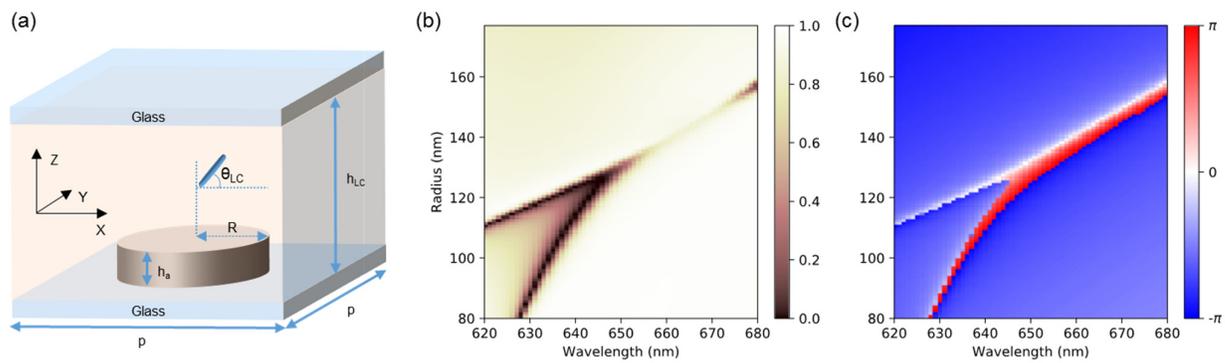

Figure 2

(a) [Phase Angle (π) vs Wavelength plot, legend: 0, 45, 90; brackets labeled 2π/3]

(b) [Field distribution schematic with E_x polarization, labeled "Glass"]

(c) [Transmission vs Wavelength (nm) plot, legend: $T_{-1}$, $T_0$, $T_{+1}$]

Figure 3

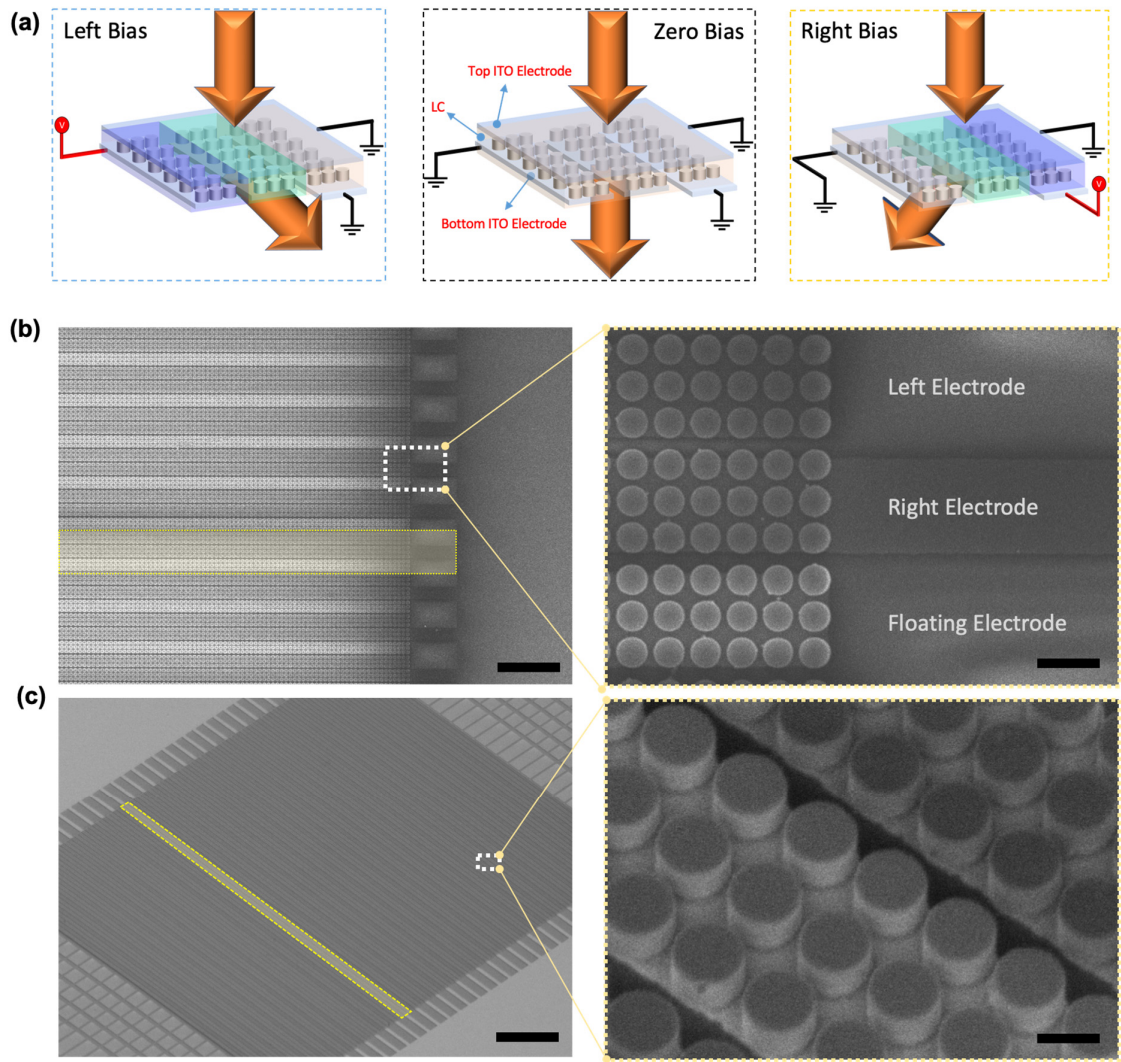

Figure 4

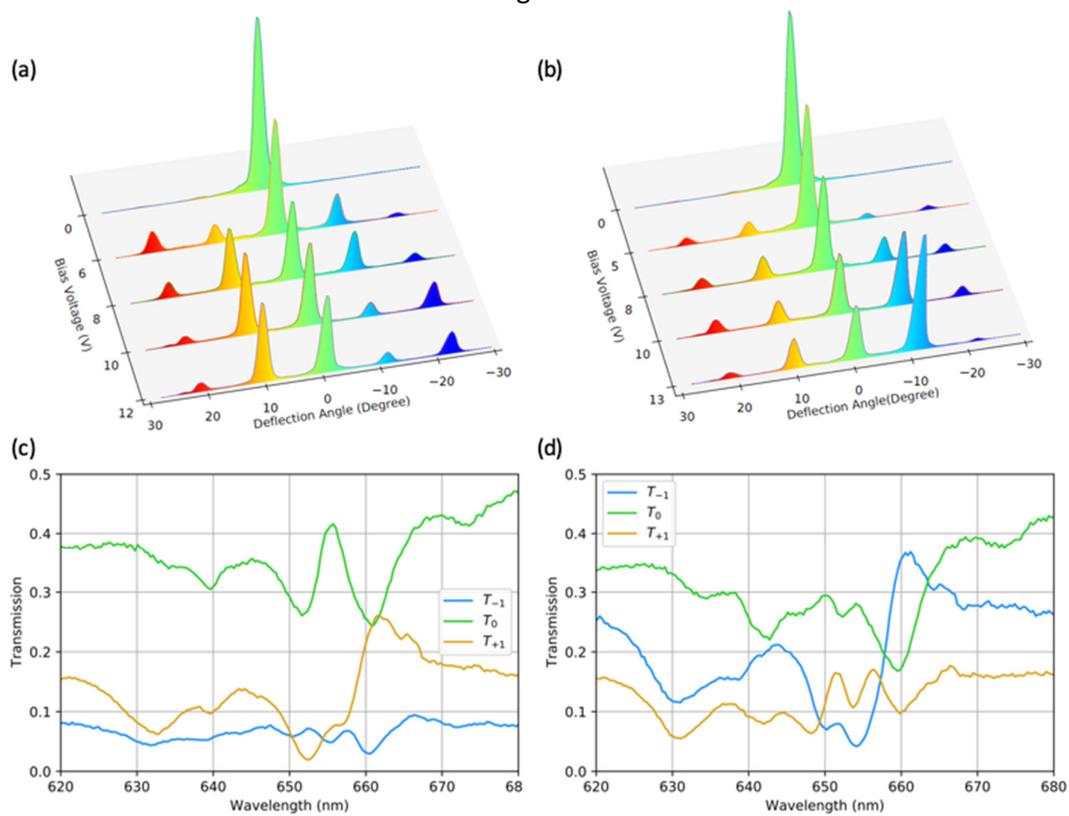

# Supplementary Materials for

## Phase-only transmissive spatial light modulator based on tunable dielectric metasurface

Shi-Qiang Li, Xuewu Xu, Rasna Maruthiyodan Veetil, Vytautas Valuckas, Ramón Paniagua-Domínguez, and Arseniy I. Kuznetsov*

Correspondence to: arseniy_kuznetsov@imre.a-star.edu.sg

**This PDF file includes:**

Materials and Methods
Figs. S1 to S6
Tables S1

**Materials and Methods**

Fabrication of the nanoantenna-based SLM device

We start with a commercial ITO-coated conductive glass slide (Latech Scientific Supply Pte. Ltd, Singapore) with ITO film thickness of 23 ± 5 nm (Figure S6a). Measured transmittance of the glass slide is above 87% in the whole visible spectral range and resistivity is ~100 Ω/□. Amorphous $TiO_2$ film is then deposited onto the ITO glass slides with Oxford Optofab 3000 ion assisted deposition system (Figure S6b). The deposition rate is 0.28 Å/s to ensure the smoothness of the films. The surface roughness after the $TiO_2$ film deposition is characterized by Vecco DI 3100 Atomic Force Microscope and the root mean square value obtained is around 1.0 nm from several random areas with size of 10 by 10 microns. Following that, a 30 nm Chromium (Cr) film is evaporated on top of the $TiO_2$ film by an electron-beam evaporator (Angstrom EcoVac), as shown in Figure S6c. Electron-beam lithography (EBL) is then performed to define the masks for the nanoantennas using Elionix ELS-7000 EBL system. A standard procedure is conducted using a negative tone electron-beam resist, Hydrogen silsesquioxane (HSQ, Fox-22, Dow-Corning Inc.), to define the mask pattern of the nanoantennas (Figure S6d&e). Following that, the pattern is transferred to the Cr layer by a reactive ion etching process in Oxford PlasmaLab 100. Processes with $Cl_2$ and $O_2$ gases are used to ensure the fidelity of the pattern transfer due to their etch selectivity of Cr over HSQ (Figure S6f). With Cr as a hard mask, formation of the $TiO_2$ nanoantennas is achieved through dry etching with $CHF_3$ gas (Figure S6g) in the same etcher. Chromium mask is then removed by immersing the sample in a chromium etchant solution (Sigma-Aldrich) for 4 minutes (Figure S6h). A second EBL process is then performed on the sample. The standard process for the positive tone electron-beam resist (ZEP520A, ZEONREX Electronic Chemicals) is used for the second layer. Electrode pattern is directly transferred from ZEP to the ITO layer by another reactive ion etching process in the same etcher. A recipe with $CH_4$ and Ar gases is used to achieve good etching selectivity of ITO over ZEP resist and the structure after etching is shown schematically in Figure S6i.

Scanning electron microscopy

Scanning electron microscopy (SEM) imaging of the nanofabricated samples is performed on Hitachi SU8200 Ultra-high resolution scanning electron microscope. Electron acceleration voltage used is 1 kV with a nominal current of 10 nA. The low voltage is used to prevent the charging effect on the samples fabricated on insulating glass substrates.

Assembly of the nanoantenna-based SLM device

ITO conductive glass slide with the same specification as the device substrate is used as the top covering slide to form the liquid crystal cell. Firstly, a layer of polyimide is spin coated on the covering slide followed by curing at 180 °C for 1 hour on a hot plate. The cured slides are then rubbed unidirectionally with a piece of soft velvet clothes to define a preferred liquid crystal alignment direction of the cell. The rubbing strength and duration are controlled by a commercially available rubbing machine (Holmarc Opto-Mechatronics - HO-IAD-BTR-03). Following that, the cover slide is pressed onto the device structure, with the spacing defined by a layer of glue (Norland optical adhesive NOA81). After the desired gap thickness between the two cover slides is reached, the glue layer is cured via ultra-violet light irradiation. The spacing and flatness between the top covering slide and the bottom substrate is precisely controlled through a home-made press equipped with a multi-point optical profilometer. Lastly, nematic liquid crystal molecules (E7 from Merck) are introduced into the gap formed between the top covering slide and the bottom substrate by capillary forces and vacuum pumping. After the infiltration the nematic liquid crystal directors are aligned parallel with the alignment direction introduced by the rubbing process on the covering slide. Figure S6j shows a schematic of the assembled device.

Setup for optical characterization and beam bending measurements.

The optical characterization setup used in the experiments has been described in details elsewhere [49]. In brief, the incident beam from a halogen lamp is linearly polarized along the shorter axis of the bottom electrodes of the SLM. A pupil diaphragm on the optical condenser is closed to its minimum to limit the angular spread of the incident light to the smallest limit (2° from the normal to the device surface). Light transmits through the device and is collected through a bright field objective lens (Nikon 50X, NA = 0.45). The back focal plane of the objective is then imaged to the entrance slit of Andor Spectrograph (SR-303i). The direction of the slit is aligned with the short axis of the electrodes so that the beam deflection along that direction is spatially resolved. Along the perpendicular direction, light is dispersed by a blazed grating so that we obtain the spectral information of the deflected beam. The image with an angular resolution of 0.38° and a spectral resolution of 0.34 nm is collected by an EMCCD (Andor Newton Series) sensor array and stored in a computer for further analysis. The active modulation of the bias voltage on the electrodes is provided by a power supply with a voltage resolution of 10 mV. (Keysight Technology U8031A Triple Output DC Power Supply)

Finite-element-method Simulation

The simulations were performed using COMSOL Multiphysics software package. The detailed procedure is a typical electromagnetic wave simulation in the frequency domain with one unit cell of the nanoantenna array as the simulation cell. An example is provided by COMSOL Inc at the following link (https://comsol.com/model/plasmonic-wire-grating-wave-optics-14705; accessed on December 29, 2018)

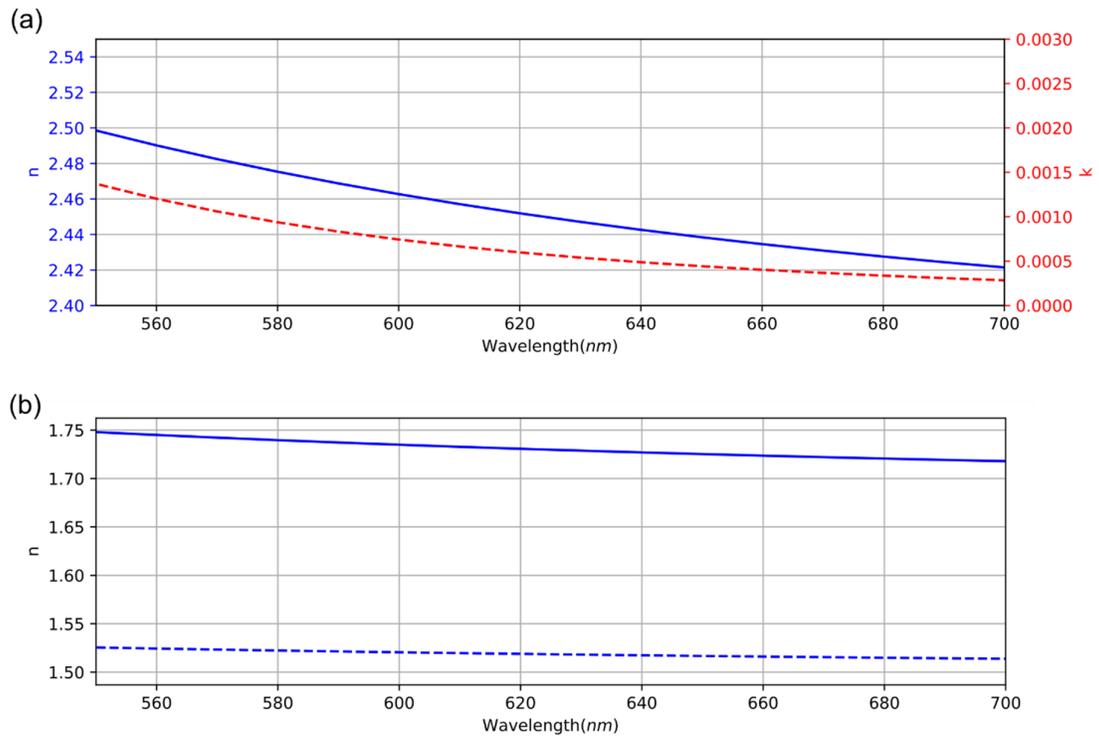

**Fig. S1.**

(a) Real and imaginary parts of refractive index (n and k) of our $TiO_2$ films extracted from ellipsometry measurements; (b) Real and imaginary part of refractive index (n and k) of E7 liquid crystal obtained from literature data.[1]

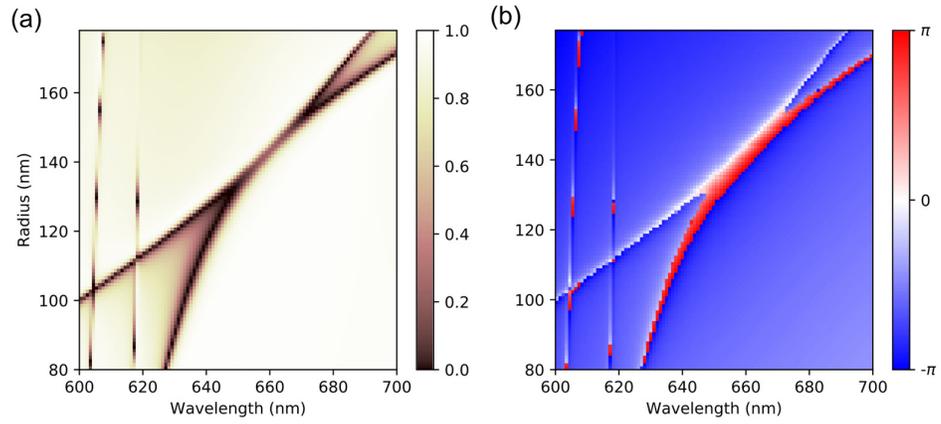

**Fig. S2.**
Numerically calculated (a) amplitude and (b) phase shift of the transmitted light for nanoantenna arrays with a height of 195 nm.

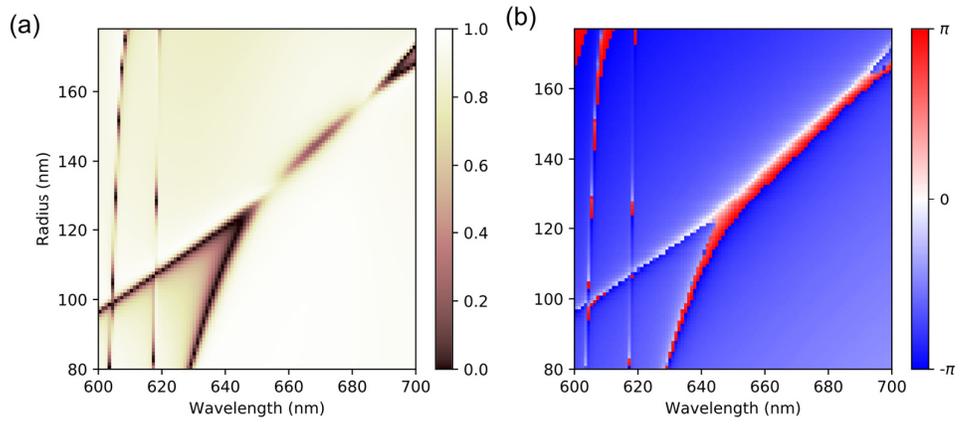

**Fig. S3.**

Numerically calculated (a) amplitude and (b) phase shift of the transmitted light for nanoantenna arrays with a height of 215 nm.

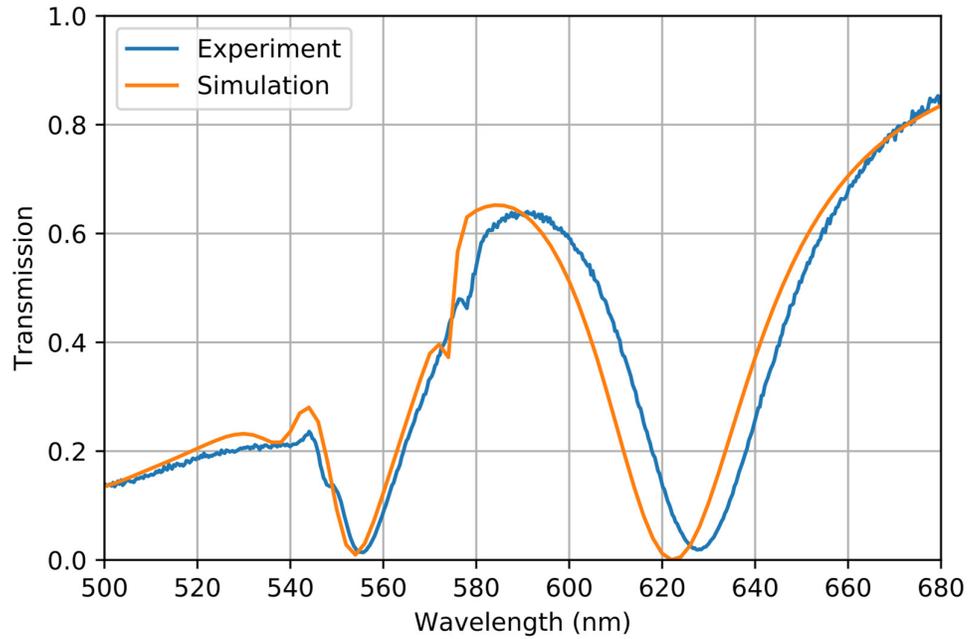

**Fig. S4.**

Optical transmission measurement of a nanoantenna array in air and its comparison to the simulations. Parameters of the simulated unit cell are similar to the nanoantenna array dimensions measured by SEM. Height of $TiO_2$ nanodisks - h = 205 nm; radius of $TiO_2$ nanodisks - R = 135 nm; period of the unit cell - p = 360 nm; 60 nm gap is included in the direction of the polarization of the incident electric field between every three unit-cells along the same direction (to simulate the introduced gap between individual electrodes). The wavevector of the incident light is normal to the surface of the glass substrate. ITO layer is ignored in the simulation as its contribution was verified to be negligible but computationally demanding.

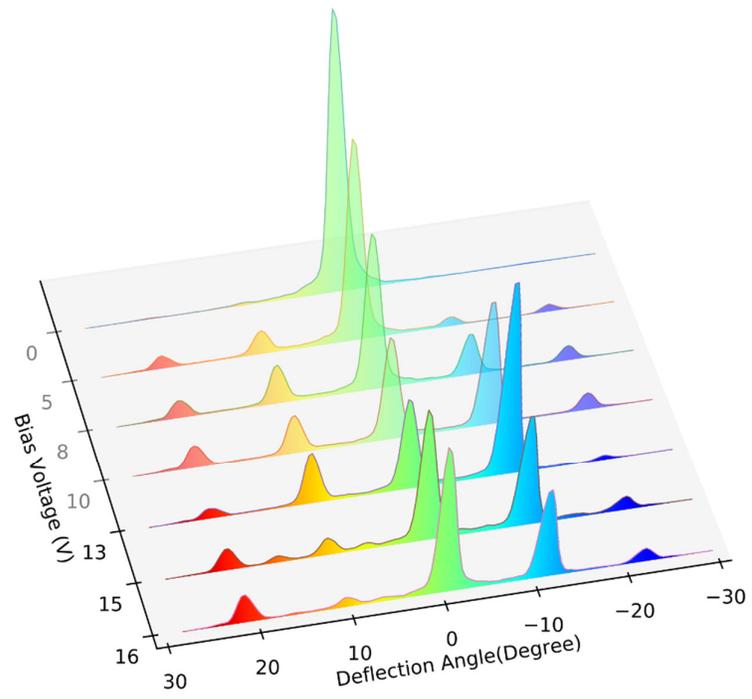

**Fig. S5.**
Efficiency of the device beyond optimum voltage. Beyond 12 volts, the efficiency drops as plotted.

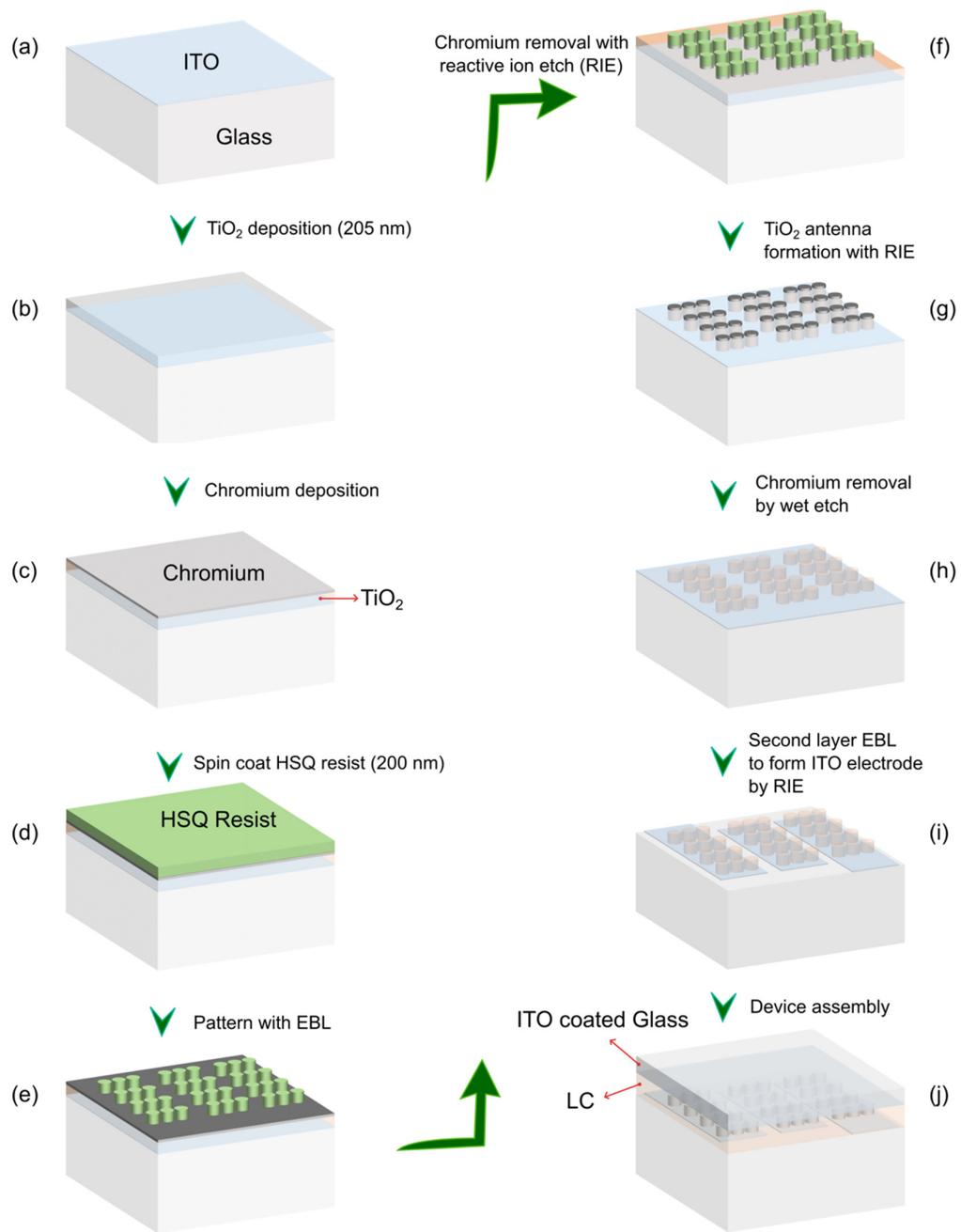

**Fig. S6.**
Process Flow of the nanoantenna-based SLM device fabrication

**Table S1.**

Light bending efficiency for 3-phase level SLM for different numbers of $TiO_2$ nanoantenns per pixel (Design "3LnP" corresponds to n nanoantennas per pixel). w: Width of each electrode; $\lambda$: Wavelength corresponding to optimum diffraction efficiency; $T_{total}$: Total light transmission; $T_0$, $T_{+1}$ and $T_{-1}$: transmission into 0, +1 and -1 diffraction orders respectively; $\theta_{-1}$: First order diffraction angle.

| Design | w (μm) | λ (nm) | $T_0$ | $T_{+1}$ | $T_{-1}$ | $T_{total}$ | $\theta_{-1}$ (°) |
|---|---|---|---|---|---|---|---|
| 3L1P | 0.36 | 666.0 | 0.002 | 0.155 | 0.288 | 0.520 | 38.07 |
| 3L2P | 0.72 | 665.5 | 0.0002 | 0.134 | 0.506 | 0.789 | 17.94 |
| 3L3P | 1.08 | 662.5 | 0.001 | 0.093 | 0.537 | 0.834 | 11.80 |
| 3L5P | 1.80 | 663.0 | 0.006 | 0.052 | 0.539 | 0.854 | 7.05 |
| 3L7P | 2.52 | 664.0 | 0.006 | 0.0520 | 0.533 | 0.869 | 5.04 |
| 3L9P | 3.24 | 664.0 | 0.007 | 0.0512 | 0.545 | 0.890 | 3.92 |